\documentclass[english,xaps,prb,twocolumn]{revtex4}
\usepackage{amsmath}
\usepackage{amssymb}
\usepackage{graphicx}

\makeatletter
\@ifundefined{textcolor}{}
{%
 \definecolor{BLACK}{gray}{0}
 \definecolor{WHITE}{gray}{1}
 \definecolor{RED}{rgb}{1,0,0}
 \definecolor{GREEN}{rgb}{0,1,0}
 \definecolor{BLUE}{rgb}{0,0,1}
 \definecolor{CYAN}{cmyk}{1,0,0,0}
 \definecolor{MAGENTA}{cmyk}{0,1,0,0}
 \definecolor{YELLOW}{cmyk}{0,0,1,0}
}

\usepackage{amsbsy}
\usepackage{mciteplus}
\usepackage{babel}

\makeatother

\usepackage{babel}
\begin{document}

\title{Imaging transition to fractional quantum Hall regime by Coulomb blockade microscopy}

\author{E. Wach}
\affiliation{AGH University of Science and Technology, Faculty of Physics and Applied Computer Science,\\
al. Mickiewicza 30, 30-059 Krak\'ow, Poland}
\author{D. P. \.Zebrowski}
\affiliation{AGH University of Science and Technology, Faculty of Physics and Applied Computer Science,\\
al. Mickiewicza 30, 30-059 Krak\'ow, Poland}
\author{B. Szafran}
\affiliation{AGH University of Science and Technology, Faculty of Physics and Applied Computer Science,\\
al. Mickiewicza 30, 30-059 Krak\'ow, Poland}

\date{\today}

\begin{abstract}
We consider electron systems in quantum dots and imaging of the confined charge density by the Coulomb blockade microscopy (CBM)
with the scanning probe technique. We apply an exact diagonalization method to study the reaction of the electron
system to the potential induced by the model potential of the probe and calculate the energy maps as functions
of the position of the probe. The charge densities derived from the energy maps are confronted to the exact charge densities.
We focus on the transition of the electron system to the fractional quantum Hall conditions in external magnetic field.
For magnetic fields corresponding to the integer fillings of the lowest Landau level the electron system 
exhibits a liquid-like reaction to the potential of the probe and the confined charge density can be quite accurately 
mapped by the CBM. 
For fractional fillings of the lowest Landau level the single-electron charge density islands nucleate 
in presence of the tip. In circular quantum dots the single-electron islands evade imaging by CBM. We demonstrate
that mapping the molecular charge densities is possible for confinement potentials of symmetry that is lower and 
consistent with the geometry of the lowest-energy charge distribution of the single-electron islands.
\end{abstract}
\maketitle
\section{Introduction}

Scanning gate microscopy (SGM) \cite{sgmr} is a technique that probes the local properties
of a semiconductor system by measurement of the current flowing through the device as a function of the position of the charged atomic force microscope tip sweeping
above the sample area.
The charged tip is capacitively coupled to the electron gas and introduces a local perturbation to the potential landscape.
The technique allows for mapping the charge flow patterns, local density of states
at the Fermi level, and/or the confined charge density.\cite{sgmr}
SGM has been applied for gated devices based on two-dimensional electron gas in III-V heterostructures,\cite{qpc0,qpc1,qpc2,qpc4,qpc8,qpc10,gild07,wff,fallahi,huefner11,pioda} on quantum dots
induced in quantum wires,\cite{blesz08,boydnano}
graphene systems \cite{graphene} and carbon nanotubes. \cite{cnt,zhang}
For open systems,\cite{qpc0,qpc1,qpc2,qpc4,qpc8,qpc10,wff} the information on the properties of the electron system
is derived from the maps of conductance variation as functions of the tip position.
For closed quantum dots -- systems that are weakly coupled to the electron reservoirs --
in conditions of the Coulomb blockade \cite{kouwen} -- the charged probe potential
switches on the current by tuning the chemical potential of the confined $N$-electron
system into the transport energy window fixed by the Fermi energies of the drain and source
electrodes.\cite{gild07,zhang,fallahi,huefner11,pioda,blesz08,boydnano}  The technique -- referred to as the Coulomb blockade microscopy \cite{qiang,mantelli}-- allows
to extract the confined charge density by the inverse problem analysis.\cite{boyd}

In this paper we consider mapping of the electron density confined in the two-dimensional quantum dot by the
Coulomb blockade microscopy in external magnetic field. We focus on the ground-state transition between the maximum density droplet (MDD) \cite{reiman}
and the ground-states at low fillings of the lowest Landau level at high magnetic fields corresponding to the fractional quantum Hall conditions.
The transition is accompanied by a pronounced step-wise increase of the electron-electron correlation effects \cite{reiman,yanourev}
with formation of electron molecules in the relative coordinates of the quantum system. The electrons in the fractional quantum Hall regime
form a phase of properties that are between the Laughlin liquid and the Wigner crystal \cite{yl,jainprl} with the latter getting stronger
with reduction of the filling factor of the lowest Landau level $\nu$. The maximum density droplet decay
at a critical magnetic field is an abrupt counterpart of the transition from the Fermi liquid to the Wigner solid at zero magnetic field that appears in
a continuous manner with the growth of the quantum dot size.\cite{guclu,balzer}

We consider extraction of the unperturbed charge density from the energy maps as functions of the tip position.
We solve the few-electron eigenproblem using the exact diagonalization approach and calculate the energy maps
for a model tip potential. We discuss the relation of the charge density as reproduced from the energy variation as a function of the tip position
to the unperturbed charge density.
We find that in circular quantum dots the reaction of the electron charge density to the tip undergoes a qualitative change at the crossing between the integer
and fractional filling factors, with a liquid-like behavior for $\nu\ge1$ and nucleation of the single-electron islands in presence of the tip for $\nu<1$.
For circular quantum dots the single-electron molecular charge distributions induced by the tip for $\nu < 1$ are not resolved by the charge densities that are reproduced
from the energy maps. However, for lower symmetry of the confinement potential
due to a defect in particular, the electron molecules appear in the laboratory frame at $\nu< 1$.
The molecular charge densities pinned by the confinement potential can be extracted by the Coulomb blockade microscopy.

The extraction of the charge densities from the Coulomb blockade microscopy
was considered in a number of papers but uniquely for one-dimensional quantum dots.\cite{qiang,mantelli,boyd,ziani}
We show below, that quasi one-dimensional systems in which the few-electron charge density is rigid, the magnetic field has no pronounced
influence on the charge density mapping with the SGM. This is in contrast to the two-dimensional
quantum dots in which the role of the magnetic field is very pronounced.
The study of two-dimensional quantum dots in the absence of the magnetic field was given in our previous paper.\cite{wachjpcm}

This paper is organized as follows: Section II explains the model and computational approach; Section III presents the results for
an ideally circular parabolic quantum dot, for the quantum dot perturbed by a presence of a charged defect, as well as for confinement
potential of a rectangular shape and flat profile near the minimum. Summary and conclusions are given in Section~IV.

\section{Theory}
The system of $N$ electrons is described by the effective-mass Hamiltonian
\begin{equation}
H=\sum_{i=1}^N  h({\bf r_i}) +\sum_{i=1}^N \sum_{j=i+1}^N \frac{e^2}{4\pi\epsilon\epsilon_0 r_{ij}}+B S_z g^*\mu_B,
\end{equation}
where  ${\bf r_i}$~is the position of the $i$-th electron and the last term introduces the spin Zeeman effect polarizing the ground-state spin $S_z$ at high
magnetic field $B$ oriented perpendicular to the plane of confinement. The two-dimensional single-electron energy operator is in form
\begin{equation}
h=\frac{(-i\hbar\nabla +e{\bf A})^2}{2m^*}+V_c({\bf r})+V_t({\bf r};{\bf r_t}),
\end{equation}
where $V_c$ and $V_t$ are the confinement and the tip potentials. We use the symmetric gauge ${\bf A}=B(-\frac{y}{2},\frac{x}{2},0)$,
for which the kinetic energy operator commutes with the operator of the $z$-component of the angular momentum  $L$,
and GaAs material parameters: the effective mass $m^*=0.067m_0$,
dielectric constant $\epsilon=12.5$, and the effective Land\'e factor $g^*=-0.44$.
In the following we consider a circular parabolic confinement potentials,
\begin{equation}
V_c({\bf r})=\frac{1}{2} m^* \omega^2 (x^2+y^2)-\frac{Ze^2}{4\pi\epsilon\epsilon_0 |{\bf r}-{\bf r_{def}}|},\label{h1e}
\end{equation}
with $\hbar\omega=2$ meV. The second term introduces the perturbation by a negatively charged defect (ionized donor, or charged acceptor)
at position ${\bf r_{def}}=(x_{def},0,z_{def})$ at a distance from the  plane of confinement, with $Z=1$ or $Z=0$ (in the absence of the defect).
The electrostatic confinement potentials have usually parabolic profiles near the minimum.\cite{bednarek}  However, for the specific gating geometry
systems of steeper profiles with flat minima can be produced.\cite{bednarek} The reaction of the system to the perturbation by the tip potential
should naturally be much stronger for flat potentials. The present discussion covers also this type of conditions. For that purpose we consider 
a non-parabolic confinement potential of a rectangular profile 
\begin{equation}
V_c({\bf r})=V_0 \left( 1 - \frac{1}{1 + \left( \frac{x}{X}\right)^{10} + \left( \frac{y}{Y}\right)^{10} } \right).
\label{rect}
\end{equation}
For strongly elongated rectangular quantum dots ($X \gg Y$) the confinement approach the quasi one-dimensional limit with
all the electrons occupying the same state of spatial quantization along the narrow side of the rectangle.
The mapping of the charge density is discussed for parameters $X=75$ nm, $Y=15$ nm.
We also consider a square shape of the potential ($X = Y = 50$ nm) for discussion of the relation between the intrinsic symmetry of the lowest-energy
spatial configuration of a few electrons and the symmetry of confinement in the context of the external perturbation.

The tip of the scanning gate microscope interacts with the electron system buried beneath the surface of the structure with
the electrostatic Coulomb potential which is screened by the deformation it induces on the two-dimensional electron gas.
The resulting effective potential of the tip -- as obtained by the Schr\"odinger-Poisson calculations \cite{kolasinskiszafran} -- is short range
and close to the Lorentz function which we use as the model potential in this work
\begin{equation}
V_t({\bf r};{\bf r_t})=\frac{V_T d^2}{(x-x_t)^2+(y-y_t)^2+d^2},
\end{equation}
where ${\bf r_t}=(x_t,y_t)$ is the position of the tip above the plane of confinement, $V_T$ and $d$ are the height and width of the Lorentzian.

The single-electron problem is solved with Gaussian basis functions $f_k$.\cite{chwiej} The $n$-th eigenfunction of the single-electron Hamiltonian is expanded in the
basis
\begin{equation} \phi_n(x,y)=\sum_{k=1}^{M\times M} c_k^{(n)} f_k(x,y),\end{equation}
with the centers of Gaussians ${\bf R_k}=(X_k, Y_k)$ placed on a rectangular mesh of $M\times M$ points,
\begin{equation}
f_k=\frac{1}{\alpha\sqrt{\pi}}\exp\left[-\frac{({\bf r}-{\bf R_k})^2}{2\alpha^2}+\frac{ieB}{2\hbar}(xY_k-yX_k) \right],
\end{equation}
where the imaginary term of the exponent accounts for the magnetic translation, and $\alpha$ is a variational parameter.
We use up to $M\times M=625$ centers spaced by a distance that is adopted variationally to the size of the
quantum dot. For the solution of the few-electron system we use the standard approach of the configuration interaction method
forming a basis of the Slater determinants of single-electron spin-orbitals with the spatial eigenfunctions of the single-electron
Hamiltonian $\phi_n$ and a fixed value of $S_z$.
For convergence of the energy with a precision of $\mu$eV we take up to $38$ single-electron spin-orbitals which for $N=4$ electrons produces basis of $73815$ Slater determinants.

The experiments using the Coulomb blockade microscopy \cite{gild07,blesz08,fallahi,huefner11} scan the surface of the devices inducing
the current flow when the chemical potential of the few-electron system ($\mu_N=E_N-E_{N-1}$) interacting with the probe fits in the
transport energy window between the Fermi energies of the source and drain. Since $\mu_1=E_1$, the energy maps as functions
of the tip position can be extracted for any $N$. Under assumption of a frozen electron charge density $n_r$
the energy of the $N$ electron system is a convolution of the tip potential and the confined electron density\cite{boyd}
\begin{equation} E_N({\bf r_t})=E_N(\infty)+\int_{-\infty}^\infty dx\int_{-\infty}^\infty dy V_t({\bf r};{\bf r_t}) n_r({\bf r}).\label{int}\end{equation}
In fact the confined electron density is deformed by the tip potential. The actual electron density -- with or without the tip --
is calculated from the few-electron wave function $\Psi$ as a quantum mechanical expectation value of the density operator
$n({\bf r})=\langle \Psi |\sum_{i=1}^{N} \delta ({\bf r}-{\bf r}_i) |\Psi \rangle$.
We treat Eq. (\ref{int}) as an integral equation for $n_r$ and solve it on a grid of points in the real space.
We focus on the fidelity of the charge density reproduced  from the integral equation $n_r$ with respect to the
charge density $n$ in the absence of the tip. The quantitative similarity between the $n_r$ and $n$ charge density maps
is evaluated using the normalized cross-correlation coefficient
\begin{equation}
\tau=\frac{1}{S}\int_S dx dy \frac{[n(x,y)-\langle n\rangle][n_r(x,y)-\langle n_r\rangle]}{\sigma_n \sigma_{n_r}},\label{tau}
\end{equation}
where $\langle n\rangle$ and $\sigma_n$ are the mean and standard deviation values of the maps on the area $S$.

\section{Results}

\subsection{Circular quantum dot}
\begin{figure}[htbp]
\begin{tabular}{c}
a) \includegraphics[width=0.32\textwidth]{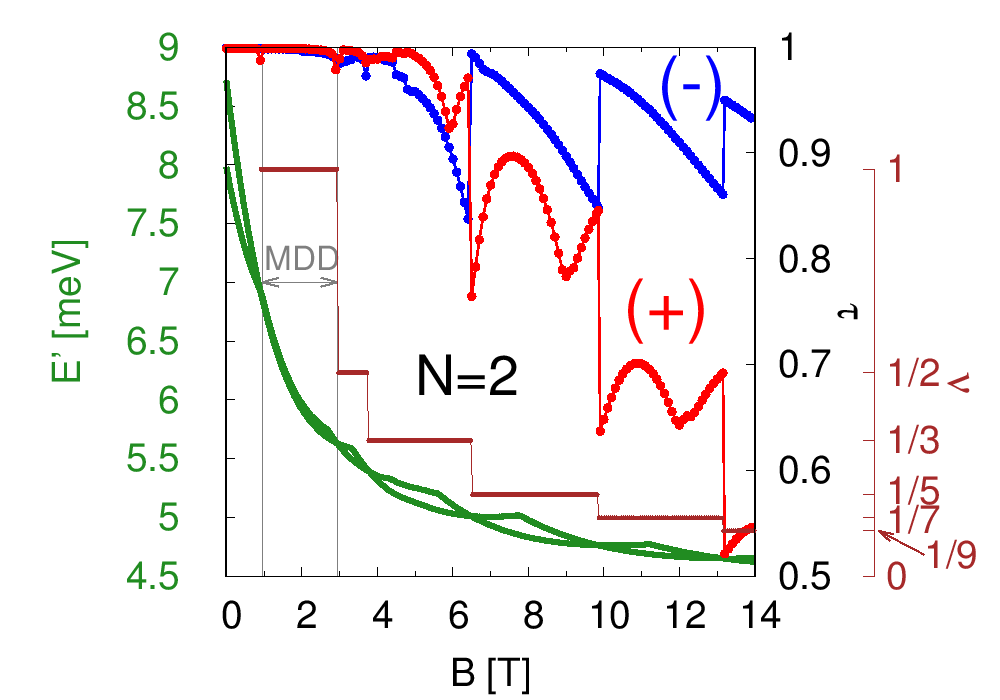} \\
b) \includegraphics[width=0.32\textwidth]{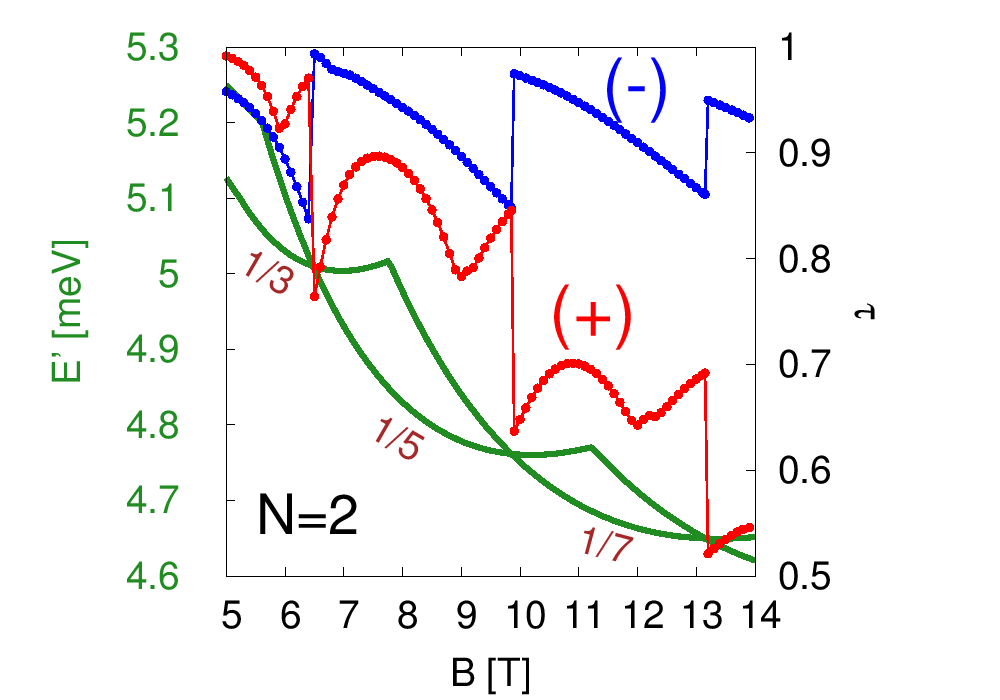} \\
\hspace{-0.9cm}c) \includegraphics[width=0.25\textwidth]{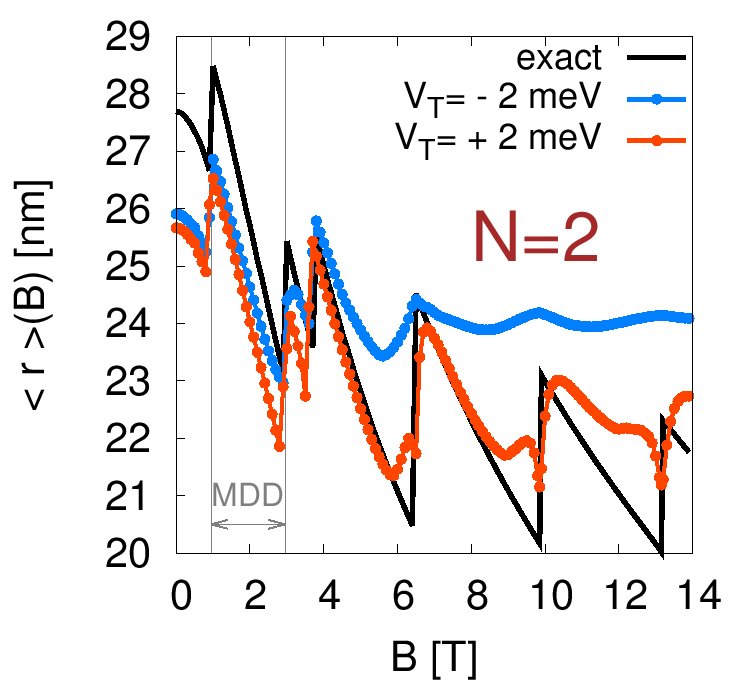}\end{tabular}
\caption{(a) Two lowest-energy levels for the electron pair calculated with respect to the lowest Landau energy level ($E'=E-N B \cdot \left(0.85 \left[ \frac{meV}{T} \right] \right)$; green lines), the filling factor $\nu$ (brown line, brown axis), and the
cross correlation $\tau$ between the unperturbed charge density and the density reproduced from the energy map as a function of the tip position [Eq. (\ref{tau})] for $V_T=\pm$ 2 meV (blue and red lines, right axis).
(b) Enlarged fragment for higher $B$. (c) The average radius of the exact charge density (black line), and the reproduced densities with blue (red) dots for $V_T=-2$ meV ($V_T=2$ meV).}
 \label{2e}
\end{figure}

\begin{figure*}[htbp]
\includegraphics[width=0.8\textwidth]{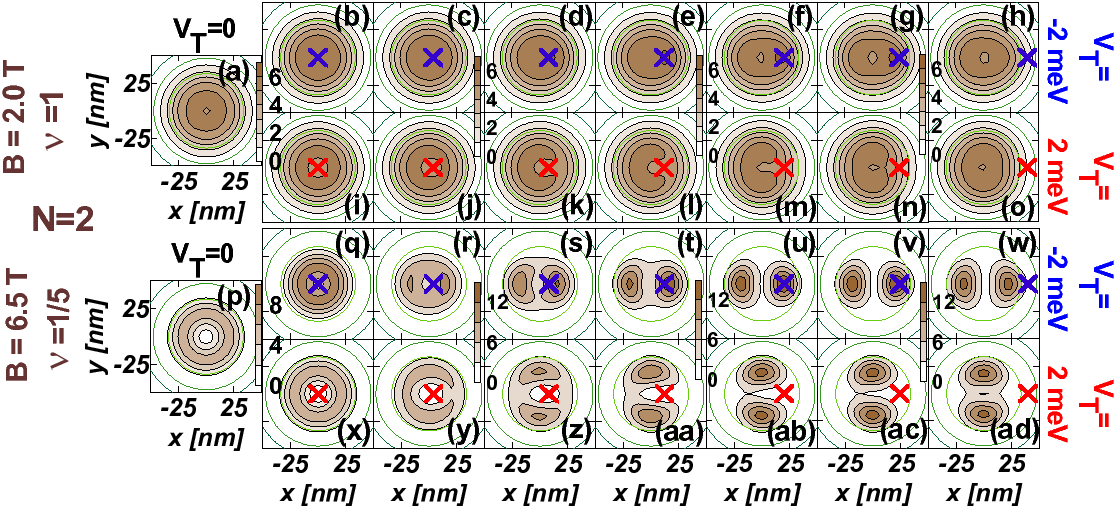}
\caption{The two-electron confined charge densities for the ground-state at $B=2$ T (MDD) (a-o) and $B=6.5$ T ($\nu=1/5$; (p-ad)) in the absence (a,p) and presence of the tip (b-o, q-ad). The position of the tip is marked with the cross. The thin green contours are equipotential lines spaced by 2 meV.}
\label{ztipem}
\end{figure*}

Figure \ref{2e} shows the two lowest-energy levels for a pair of electrons in a circular QD, with the ground-state angular momentum transitions driven by the external magnetic field.
For spin-polarized systems the value of the filling factor $\nu$ is related to the orbital angular momentum of the electron system according to formula \cite{girvin} $\nu=-\frac{N(N-1)}{2L}$.
For confinement potentials deviating from the rotational symmetry we evaluate $\nu$ using the same formula but with average values of the angular momentum operator replacing the quantum number $L$.
In Fig. \ref{2e} the maximum density droplet state $\nu=1$ ($L=-1$) appears for magnetic field between 1 T and 3 T. Above 3 T, the system enters into the fractional filling regime.
In Figure \ref{ztipem} we plotted the unperturbed [Fig. \ref{ztipem}(a,p)] charge density and the charge density as obtained
for the repulsive or the attractive tip potential $V_T=\pm 2$ meV [Fig. \ref{ztipem}(b-o,q-ad)]. For illustration in Fig. \ref{ztipem} we chose the magnetic field
of $B=2$ T for which the ground state corresponds to the filling factor of $\nu=1$ (maximum density droplet, Fig. \ref{ztipem}(a)),
and the ground-state at $B=6.5$ T for the fractional filling $\nu=1/5$ (ring-like charge density distribution characteristic
to the high angular momentum ground states, Fig. \ref{ztipem}(p)).
For the ground state with $\nu=1$ the tip produces elliptical deformation of the charge density when placed off the center of the quantum dot [Fig. \ref{ztipem}(c-h,j-o)].
A very different reaction is observed for the fractional filling factors: the tip when taken away from the center of the quantum dot induces the nucleation of a molecular charge density with the single-electron islands,
with one electron following the tip for  $V_T=-2$ meV [Fig. \ref{ztipem}(r-w)]. For the repulsive tip [$V_T=+2$ meV] the electron molecule acquires a perpendicular orientation [Fig. \ref{ztipem}(y-ad)].

\begin{figure}[htbp]
\includegraphics[width=0.5\textwidth]{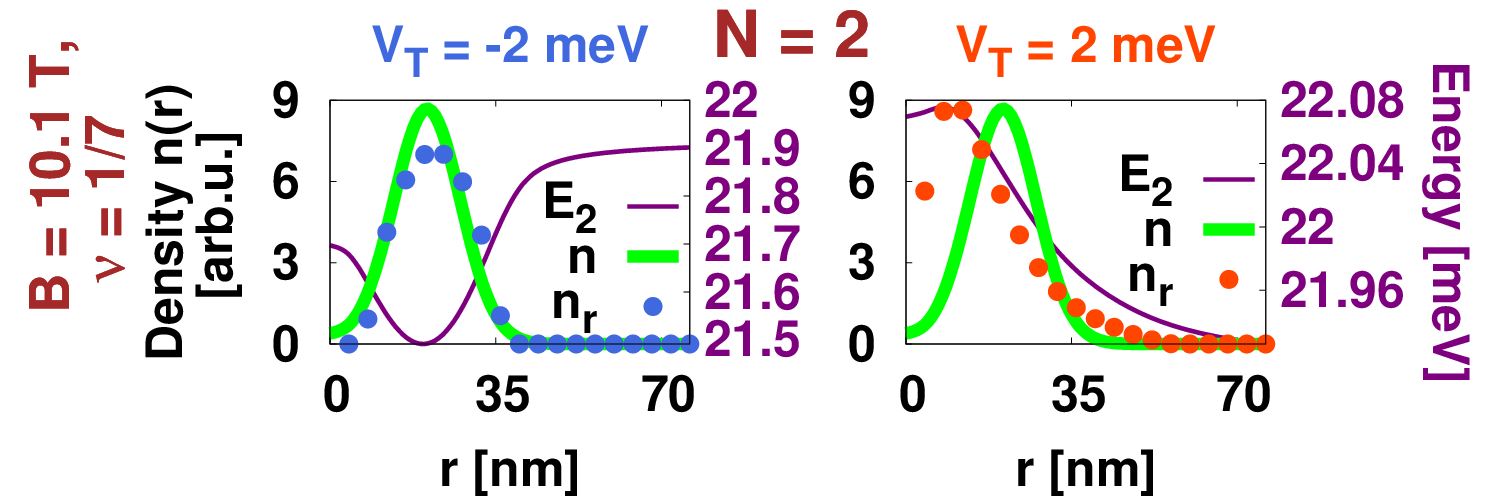}
\caption{The exact charge density (green line) and the reproduced density (dots) for $N=2$ with $B=10.1$ T and $\nu=1/7$. The purple line indicates
the energy of the system as a function of the tip position. \label{2m}}
\end{figure}

In the circular quantum dots the maps of the energy of the system as functions of the tip position above the plane of confinement reproduce the rotational symmetry of the unperturbed
charge density.
For the negative tip potential the minimum of the energy (purple line in Fig. \ref{2m}) usually coincides with the maximum of the unperturbed charge density (green lines in Fig. \ref{2m}) but the reproduced
charge density (blue dots in Fig. \ref{2m}) is less strongly localized than the unperturbed charge density.
On the other hand for the repulsive tip [Fig. \ref{2m} for $V_T=2$ meV] the maximum of the energy is shifted toward the centers of the dot and a similar
shift is acquired by the reproduced charge density. The charge density reproduced for the repulsive tip overestimates the range of the tail of the density
for large values of $r$ [Fig. \ref{2m} for $V_T=2$ meV].
The exact density for a given $\nu$ becomes more strongly localized for increasing  magnetic
field (black line in Fig. \ref{2e}(c))  which increases the interaction energy of the system and drives the transitions to lower filling factors (larger angular momentum).
Each transition is accompanied  with an increase of the radius of the charge density ring and a drop of the interaction energy.
The radii reproduced by the attractive tip in the fractional filling regime overestimate the value for the unperturbed system [Fig. \ref{2e}(c)].
The reason for this behavior is the electron density following the local minimum of
the potential energy formed under the tip when it leaves the central area of the QD [see Fig. \ref{ztipem}(c-h,r-w)].
Moreover, at low $\nu$ the radius of the reproduced density for the attractive tip becomes weakly dependent on the magnetic field  [Fig. \ref{2e}(c)].
For the repulsive tip the dependence of the average radius on the magnetic field is closer to the unperturbed one which is a result of cancellation of two opposite deviations
of the reproduced charge density: its shift to the center of the dot [Fig. \ref{2m} for $V_T=+2$ meV] and its tail for large $r$ which is longer
than the one of the unperturbed charge density.

In Figure \ref{2e}(a,b) with the blue and red lines we plotted the correlation factor $\tau$ [Eq. (\ref{tau})] between the maps of the exact unperturbed charge density and the one calculated
from Eq. (\ref{int}).
We can see that {\it i)} the correlation factor is nearly 1 for the low magnetic field including the maximum density droplet. The value of $\tau$ drops for $\nu<1$,  {\it ii)} for the fractional filling of the lowest Landau level the charge density mapping with
the attractive
tip reproduces in a closer detail the unperturbed charge density, {\it  iii)} at the ground-state angular momentum transitions opposite jumps of the correlation factors
for the attractive and repulsive tips are found.

The feature {\it i)} is a result of a weak reaction of the charge density to the tip for low magnetic fields, including the MDD ground-state. The reaction is typical to the incompressible electron liquid [Fig. \ref{ztipem}(b-o)].
Independent of the sign of the perturbation induced by the tip the maximal density of the charge stays unchanged [Fig. \ref{ztipem}(b-o)].
For the ground-states at fractional fillings, when the tip is introduced to the system the local maxima of the charge density exceed
the values characteristic to the unperturbed density [Fig. \ref{ztipem}(q-ad)], which is a signature of the compressibility of the ground state.
Moreover, the states formed after the MDD decay exhibit reaction to the tip which is characteristic of an electron solid: formation
of the single-electron islands.
For the fractional filling factor the electrons are strongly correlated and a molecular charge density distribution is formed in the internal coordinates of the system.
The external perturbation pins the electron molecule in the laboratory frame at a fixed angle. Summarizing, the reaction of the charge density to the
external perturbation introduced by the tip is much stronger for fractional fillings than for $\nu\geq 1$. Therefore, the assumption applied in the formula (\ref{int}) that the charge density
is unchanged when the tip is introduced to the system is less valid for the fractional fillings.

The observation {\it ii)} is in agreement with the result of Fig. \ref{2m}.
The value of the magnetic field $B=10.1$ T  chosen for illustration given by Fig. \ref{2m} corresponds to the radius
of the reproduced charge density ($V_T=+2$ meV) equal to the radius of the exact charge density in the absence of the tip [Fig. \ref{2e}(c)]. The radius of the charge density reproduced
by the tip for $V_T=-2$ meV is distinctly larger at $10.1$ T.
The shape of the reproduced charge density for the repulsive tip deviates more strongly from the exact charge density
than the one obtained with the attractive tip in spite of the fact that the radius of the confined charge as obtained with the repulsive tip is closer to the exact one [see Fig. \ref{2e}(c)].

The observation {\it iii)} i.e., opposite abrupt changes of the correlation factor for both signs of the tip potential at the ground-state angular momentum transitions is related to the character of the deviation of the reproduced
densities from the exact ones. The maximum of the reproduced density for the repulsive tip is shifted towards the center of the dot with respect to the exact one, and the size of the charge density island
is overestimated by the attractive tip [Fig. \ref{2e}(c) and Fig. \ref{2m}].
 At the ground-state transition the size of the unperturbed charge density jumps up (see the black curve in Fig. \ref{2e}(c)), which lowers (increases) the correlation for $V_T=+2$ meV ($V_T=-2$ meV).

\begin{figure}[htbp]
\begin{tabular}{l}
a) \includegraphics[scale=0.305]{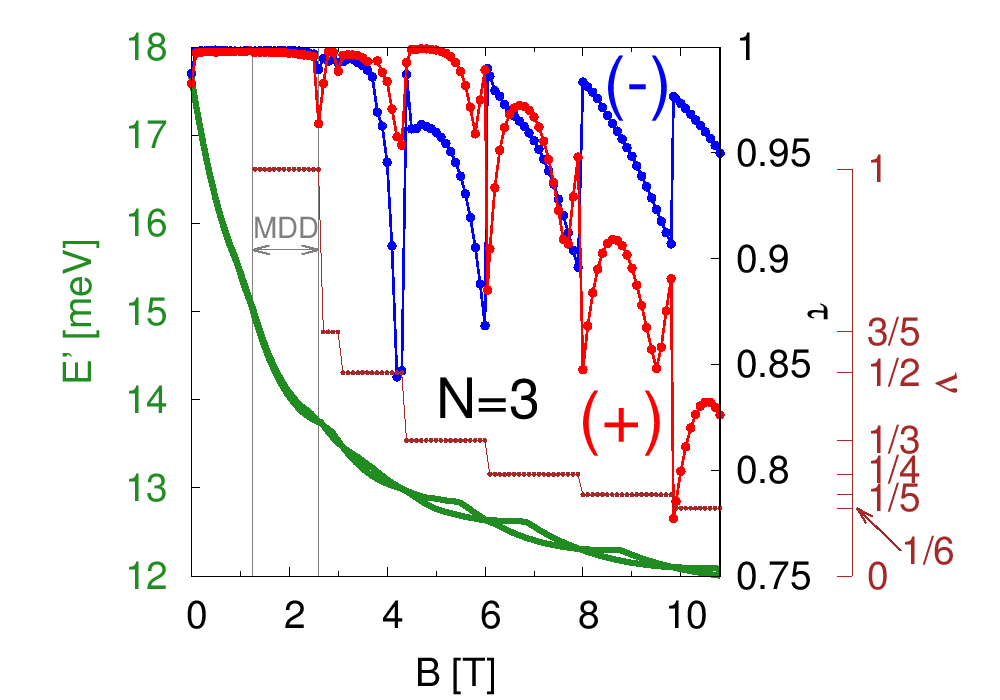}   \\
b) \includegraphics[scale=0.305]{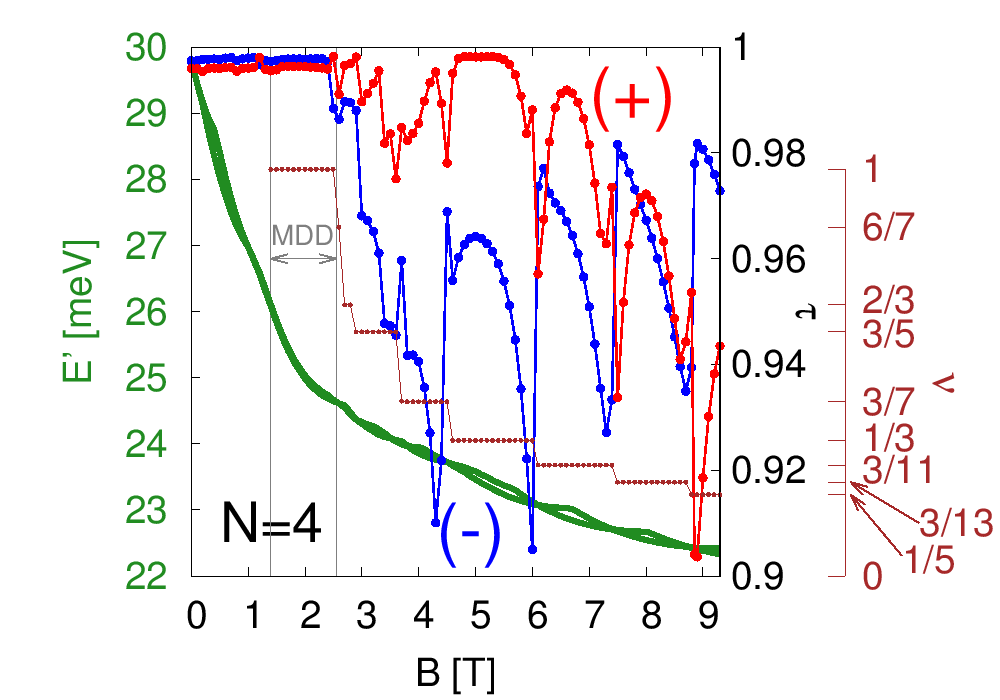}  \end{tabular}
\caption{Same as Fig. \ref{2e}(a) only for $N=3$ (a) and $N=4$ (b) electrons.\label{c34}}
\end{figure}

For larger $N=3$ and 4 [Fig. \ref{c34}] the system behaves in a qualitatively similar way to the one discussed in detail for the electron pair. The correlation factor $\tau$ is close to 1 for magnetic fields below the MDD breakdown.
The factor $\tau$ exhibits opposite jumps at the ground-state angular momentum transitions driven by the external magnetic field for the attractive and the repulsive tip potentials.
 In general, the value of $\tau$ is closer to 1 for larger $N$ -- which is a result of a stronger screening of the tip potential for larger number of the electrons.

\subsection{Circular QD with an external perturbation}
\begin{figure}[htbp]
\begin{tabular}{l}
a) \includegraphics[scale=0.3]{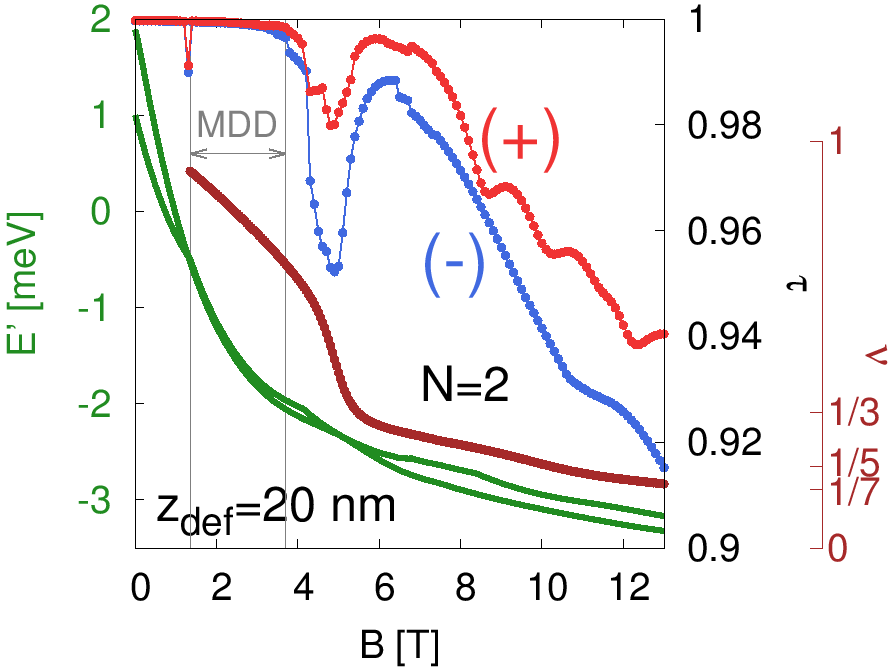} \\
b) \includegraphics[scale=0.3]{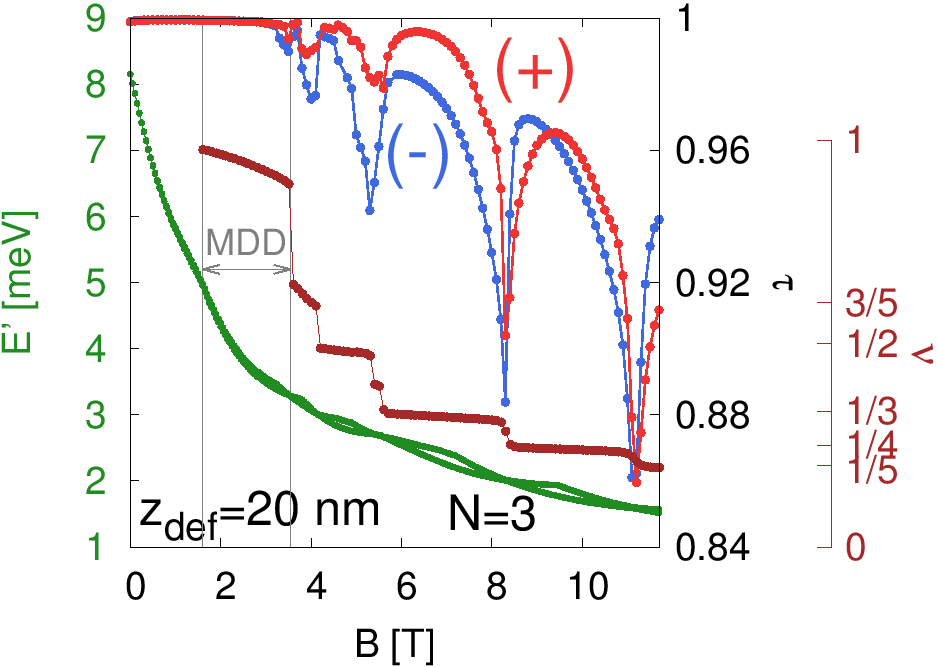} \\
c) \includegraphics[scale=0.3]{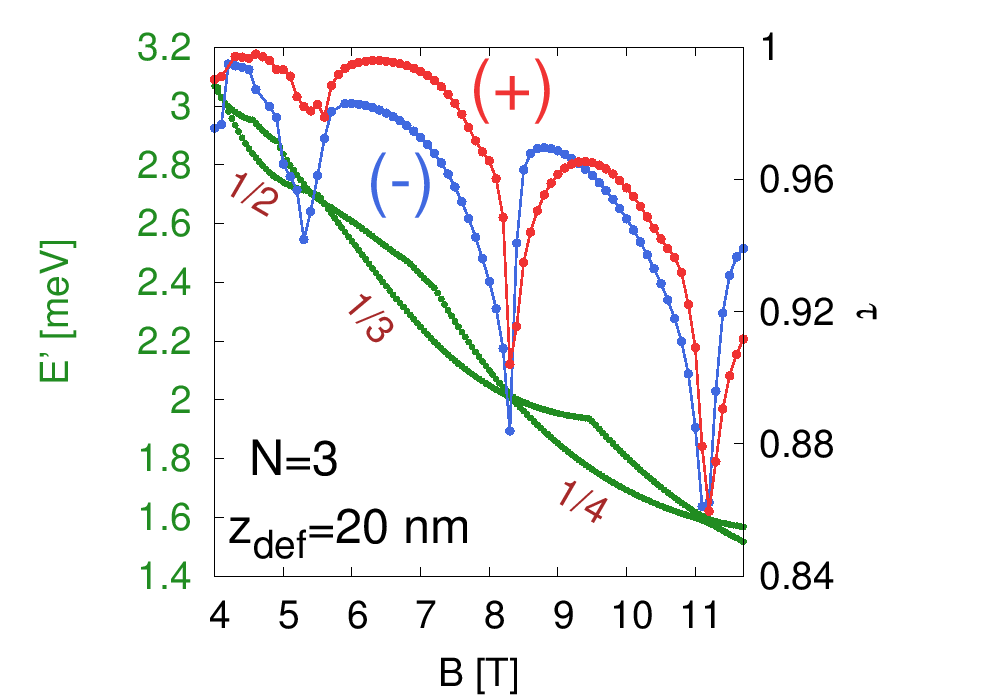}
\end{tabular}
\caption{(a) The two lowest-energy levels for the electron pair in a circular quantum dot with a charged defect above position $x_{def}=20$ nm, \mbox{$y_{def}=0$} at a distance of \mbox{$z_{def}=20$ nm} from the plane of confinement. Brown right axis refers to the filling factor $\nu$. Blue (red) dots correspond to the cross-correlation factor $\tau$ in case of $V_T=-2$ meV ($V_T=2$ meV). (b) The same as (a) but for $N=3$ electrons. (c) Enlarged fragment for higher $B$ ($N=3$).} \label{df}
\end{figure}

\begin{figure}[htbp]
    \includegraphics[scale=0.32]{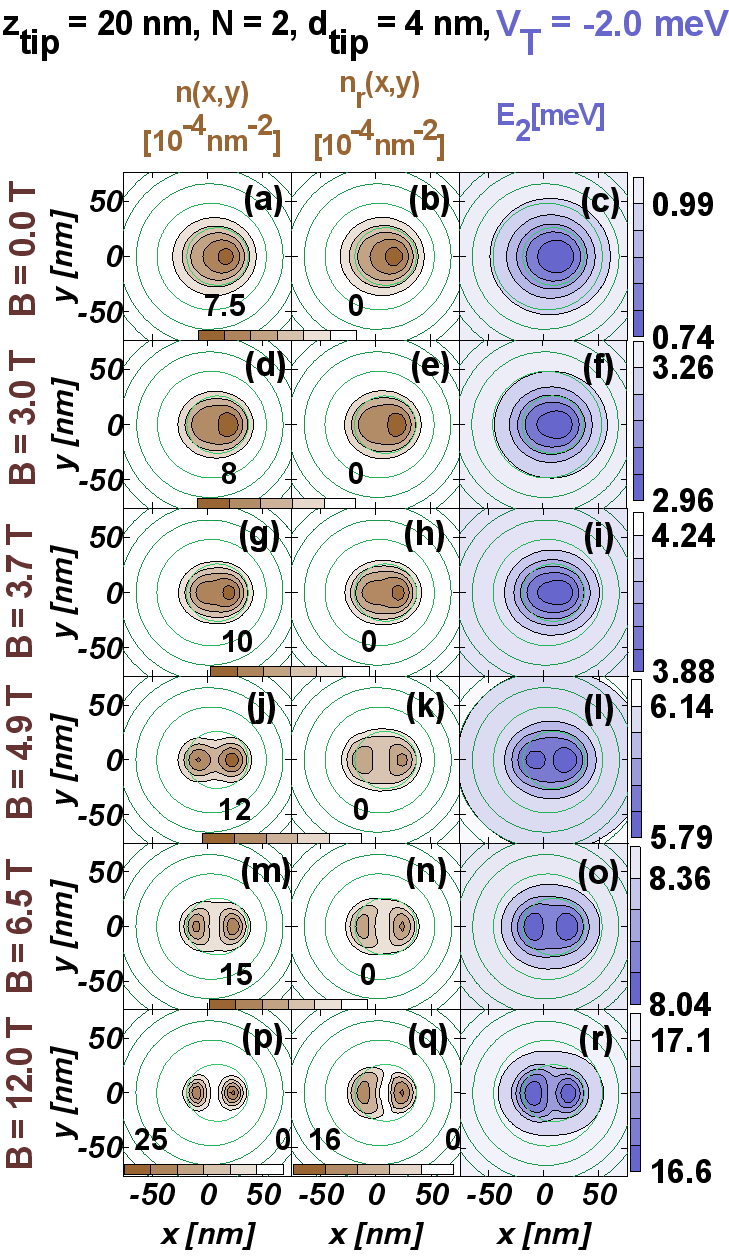}
    \caption{The charge density in the absence of the tip (left column of plots) for $N=2$ electrons for different magnetic fields; central column: the charge density reproduced from the energy maps as functions of the tip position (right column)
for \mbox{$V_T=-2$ meV}. The thin green contours are equipotential lines for the confinement potential spaced by 4 meV.}
	\label{2e_cdc}
\end{figure}

The reproduced charge densities for the circular dot are rotationally invariant as the energy maps as functions of the tip positions are.
Observation of the single-electron islands in the reproduced charge density $n_r$ even in the fractional filling conditions -- when they are
actually formed in the density $n$ in presence of the tip -- is excluded by this symmetry.
However, in real potentials a deviation from the strict symmetry is inevitable.
Let us consider a singly charged defect ($Z=1$ in Eq. \ref{h1e}) -- a ionized donor center for instance -- localized
at a distance of $z_{def}=20$ nm above the plane of confinement and placed $x_{def}=20$ nm off the center of the dot.
For two electrons the perturbation introduced by the external charge to the energy spectrum [Fig. \ref{df}(a)] is very strong.
The average angular momentum and thus the filling factor become a continuous function of $B$ with a pronounced splitting
of the ground-state and the first excited energy levels instead of the ground-state angular momentum transitions.
At low magnetic field up to the state equivalent to the MDD ($B\simeq 3.7$ T) the two-electron charge density $n(x,y)$  [Fig. \ref{2e_cdc}(a,d,g)] exhibits a shift of the maximum to the position of the charged defect.
The charge density reproduced from the energy map
very well reproduces the exact density [Fig. \ref{2e_cdc}(b,e,h)].  For $B=4.9$ T the value of $\nu$ in Fig. \ref{df}(a) drops
to about $1/3$ and the charge density nucleates in two single-electron islands [Fig. \ref{2e_cdc}(j)]. The energy map and next the reproduced charge density $n_r$ [Fig. \ref{2e_cdc}(k)] resolves the two maxima
of the original charge density $n$, although the localization degree of the single-electron maxima is underestimated by $n_r$.

\begin{figure}[htbp]
    \includegraphics[scale=0.32]{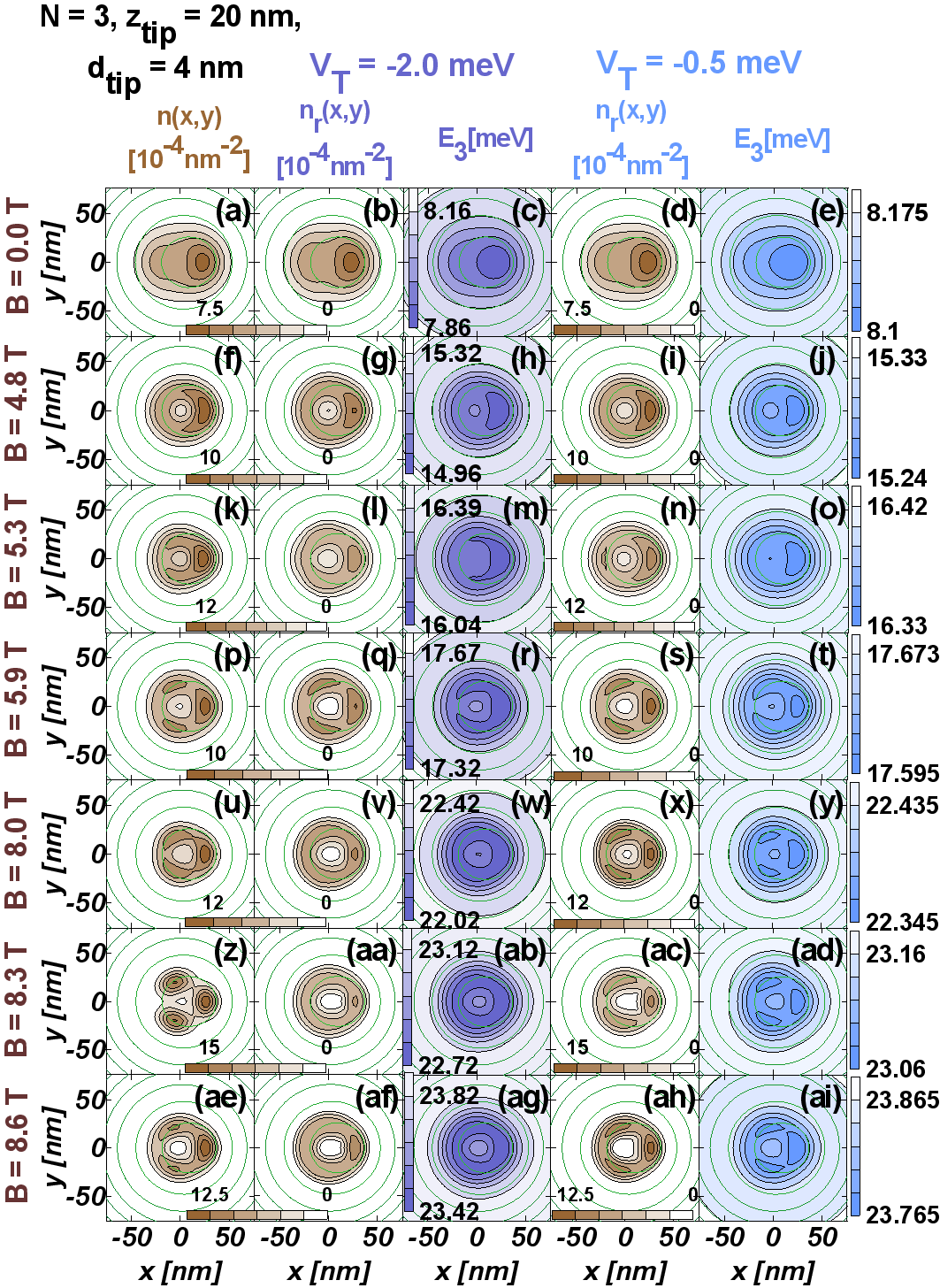}
    \caption{The first three columns from the left -- same as Fig.~\ref{2e_cdc} only for $N=3$ electrons.  The last two columns show the reproduced charge density
and the energy map obtained for $V_T=-0.5$ meV.}
	\label{3e_cdc}
\end{figure}

For $N=3$ electrons the perturbation introduced by the defect has a lower impact on the energy spectra and eigenfunctions.
The values of $\nu$ tend to reproduce the step-like dependence of the ideally circular quantum dot [see Fig. \ref{df}(b,c)]
and in the ground-state narrow avoided crossings involving states with nearly definite angular momenta are found.
All the charge densities displayed in Fig. \ref{3e_cdc} in the first column on the left-hand side as
obtained in the absence of the tip exhibit a local maximum under the position of the defect. The nucleation of two other single-electron maxima appears only for $\nu<1$
but with a strength that varies on the magnetic field scale. As a general rule, the three-electron molecule
appears in the most pronounced manner near the ground-state avoided crossings which replace the angular momentum transitions for a perfectly circular QD
in presence of the charged defect that perturbs the symmetry of the confinement potential [cf. the $n(x,y)$ plot in Fig. \ref{3e_cdc} for $B=8$ T, $8.3$ T and $8.6$ T, with
the ground-state avoided crossing at $8.3$ T -- see Fig. \ref{df}(c)].
When the tip is applied, the electron system is frustrated with respect to the source of the pinning: the defect or the tip.
For $V_T=-2$ meV we find that in general the energy maps [Fig. \ref{3e_cdc}, central column] resolve only a single extremum
under position of the defect, where both the impurity and the tip superpose to form a stronger perturbation.
For the presented set of plots [Fig. \ref{3e_cdc}, central column], only the one for $B=5.9$ T resolves the molecular structure of the charge density with three single-electron maxima [Fig. \ref{3e_cdc}(q)].
Nevertheless, for weaker perturbation introduced by the tip -- $V_T=-0.5$ meV -- the three maxima of the charge density are generally well resolved.
In the experimental conditions at fractional filling of the lowest Landau level for resolution by the Coulomb microscopy
the  electron molecule should be pinned strongly by the confinement potential to prevent reorientation
by the tip.

\subsection{Square quantum dot}

\begin{figure}[htbp]
\includegraphics[scale=0.43]{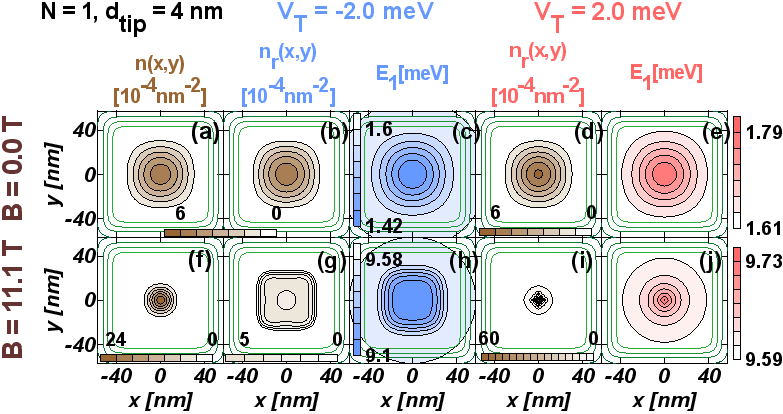}
\caption{Single electron charge densities -- exact ($n$, first column) and reproduced $n_r$ from the energy map $E_1$ for the square quantum dot. The thin green contours are equipotential lines for the confinement potential spaced by 20 meV.
The cross correlation factors are: $\tau=0.9992$ for maps (a,b), 
$\tau=0.9993$ for (a,d),  $\tau=0.5341$ for (f,g) and  $\tau=0.6419$ for (f,i).}
\label{1esquare}
\end{figure}

\begin{figure}[htbp]
\includegraphics[scale=0.38]{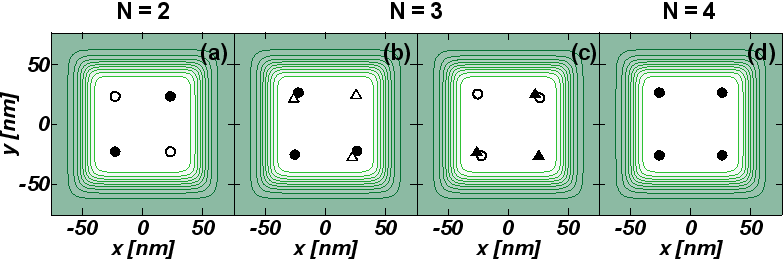}
\caption{Classical lowest energy configurations of point charges for the potential of a square quantum dot. For $N=2$ there are two equivalent configurations that are marked with different symbols (panel (a)). For $N=3$ four equivalent configurations appear (panels (b,c)). The thin green contours are equipotential lines for the confinement potential spaced by 10 meV.}
\label{class}
\end{figure}

\begin{figure}[htbp]
a) \includegraphics[scale=0.34]{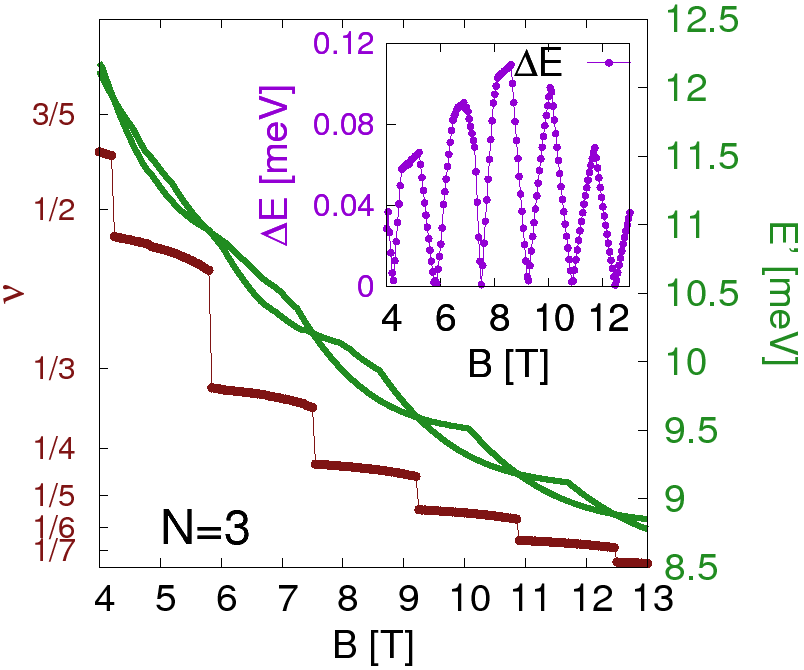} \\
b) \includegraphics[scale=0.34]{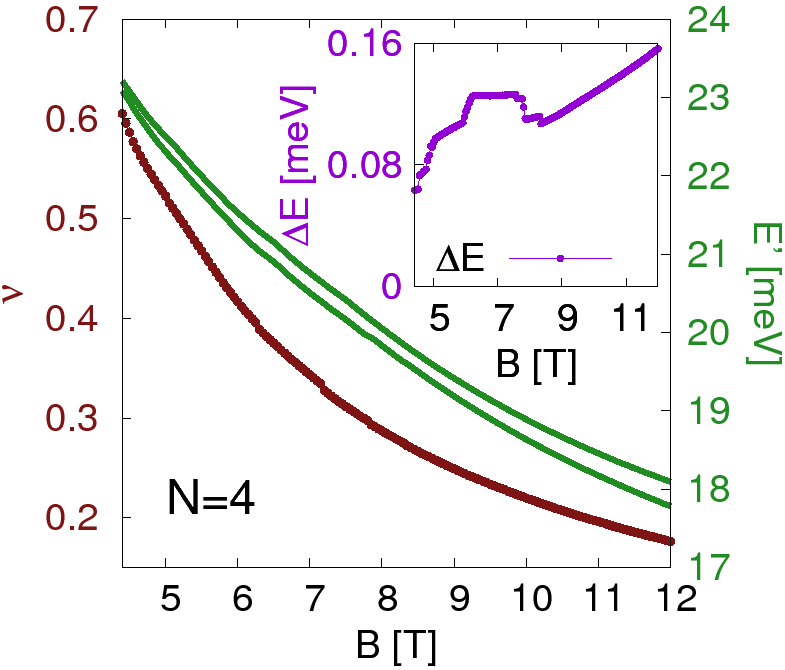}
\caption{(a) Two lowest-energy levels for $N=3$ electrons in a square quantum dot (green lines); the filling factor $\nu$ (brown line). The inset presents the energy spacing between the two lowest-energy levels. (b) The same as (a) but for $N=4$.}
\label{3e4es}
\end{figure}

\begin{figure}[htbp]
\includegraphics[scale=0.5]{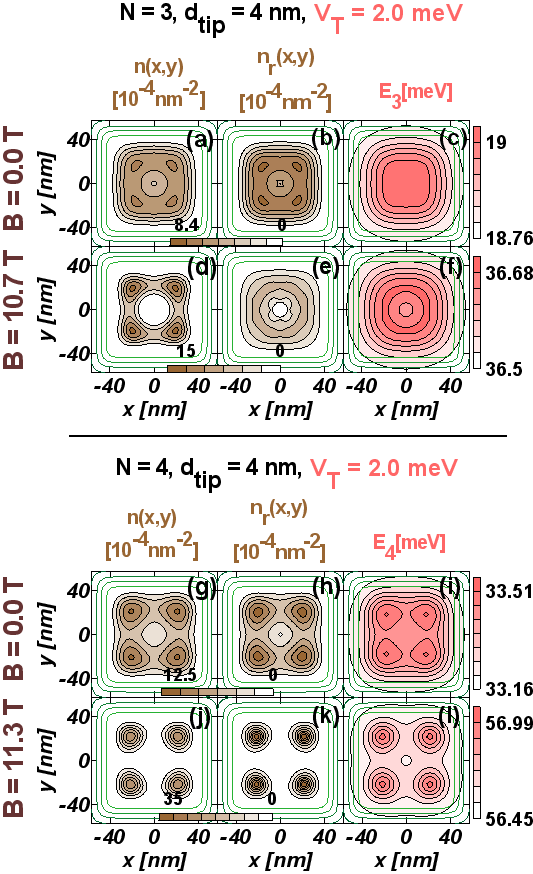}
\caption{$N=3$ and $N=4$ electron charge densities -- exact ($n$) and reproduced $n_r$ from the energy map $E_N$. 
The cross correlation factors are: $\tau=0.9970$ for (a-b), $\tau=0.8781$ for (d-e),  $\tau=0.9986$ for (g-h), and  $\tau=0.9911$ for (j-k).}
\label{3e4ed}
\end{figure}

In this section we consider confinement potential which
is flatter near the minimum and with a four-fold symmetry square shape.
The results for the exact and reproduced densities for a single confined electron are displayed in Fig. \ref{1esquare}.
At low magnetic fields the charge density is reproduced from the energy map quite accurately [Fig. \ref{1esquare} for $B=0$].
At high magnetic field, when many lowest-Landau level orbits fit in the flat quantum dot area \cite{uwage} the tip freely moves the electron inside the quantum dot.
For the attractive tip the reproduced electron density is uniformly spread inside
the quantum dot area [Fig. \ref{1esquare}(g)]. On the other hand the density reproduced for the repulsive defect has a delta-like distribution [Fig.~\ref{1esquare}(i)].
For both attractive and repulsive tip potentials the correlation coefficient between the exact and the reproduced charge density maps becomes very low at high magnetic field.

The systems of classical point charges (Fig. \ref{class}) for $N=2$ and $N=3$ electrons in the square quantum dot possess more than one configurations
of the lowest potential energy: with electrons occupying diagonal corners of the square for $N=2$ and with one of the corners unoccupied
for $N=3$. In these conditions of a classical degeneracy \cite{wachjpcm} the system undergoes parity symmetry transformations
in external magnetic field which are counterparts of the angular momentum transitions for the circular quantum dots.
The high-magnetic field spectrum for $N=3$ electrons is given in Fig. \ref{3e4es}(a). For the four-electron system the classical configuration is unique [Fig. \ref{class}(d)]
and no symmetry transitions are observed in the ground-state [Fig. \ref{3e4es}(b)]. The four-electron spectrum in the square quantum dot resembles in this respect the
two-electron spectrum for the perturbed circular dot [Fig. \ref{df}(a)]. At low magnetic fields the systems of a few-electrons exhibit local maxima near the corners
of the quantum dot [Fig. \ref{3e4ed}]. The four electron system nucleates at low $\nu$ into single-electron islands [Fig. \ref{3e4ed}(j) for $B=11.3$ T], similarly
as for the two-electrons in the perturbed circular dot [Fig. \ref{2e_cdc}]. The single-electron islands are very well resolved by the reproduced charge density.
For three [Fig. \ref{3e4ed}(d,e)] and two electrons (not shown) the density exhibits maxima at the corners of the potential, which are not resolved
by the charge density mapping at high field.

\begin{figure}[htbp]
\includegraphics[scale=0.4]{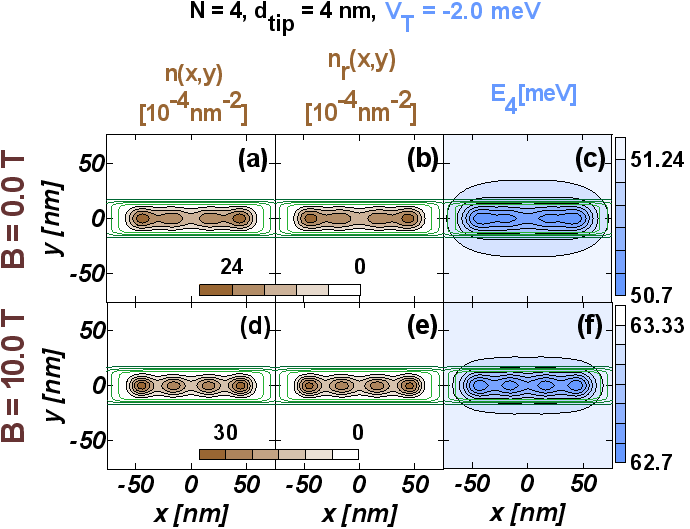}
\caption{$N=4$ electron charge densities in the rectangular quantum dot with nearly 1D confinement.  The cross correlation 
factors are  $\tau=0.9996$ for (a-b) and  $\tau=0.9994$ for (d-e). The thin green contours are equipotential lines for the confinement potential spaced by 20 meV.}
\label{1d}
\end{figure}

We also studied the rectangular potentials tending to the limit of 1D confinement \cite{zhang,qiang,boydnano,mantelli,boyd,blesz08,ziani}. In these conditions the lowest-energy configuration for any number of electrons
is unique,  and the single-electron charge islands which nucleate at high magnetic field [cf. Fig. \ref{1d}] are very stable against perturbation by the tip.
High correlation factors are obtained between the unperturbed and reproduced charge densities both at low and high magnetic field.

\section{Summary and conclusions}

We have studied the reaction of a few-electron system to the external perturbation introduced by the tip of a scanning gate
microscope using a short-range potential model and the configuration interaction approach in the context of the Coulomb blockade microscopy.
The study covered the passage of the system from low magnetic fields to the quantum Hall regime of the fractional fillings of the lowest Landau level.
We calculated the maps of the energy variation as functions of the tip position and solved the inverse problem
for the charge density. 

We studied the quantitative similarity of the unperturbed charge density to the reproduced one and
found that, generally the charge density is quite accurately described for magnetic fields below the maximum density droplet decay.
In the magnetic fields above the maximum density droplet decay, the electron system is strongly correlated and molecular
densities appear in presence of the external perturbation. In highly symmetric potentials the molecular densities evade mapping by the Coulomb blockade microscopy,
rotating with the shifts of the tip potential. Nevertheless, for systems with strongly perturbed symmetry of the confinement potential
the nucleation of the molecular density is possible. The presence of the high symmetry of the confinement potential is related to the
ground-state transitions involving cusps in the charging lines as functions of the external magnetic field. The deviations of the confinement potential
replaces the ground-state energy level crossings into avoided crossings. The wider the avoided crossing the more pronounced is the molecular form of the charge density.
In order to resolve the density the perturbation of the energy landscape by the tip should not prevail the intrinsic asymmetry of the confinement potential.
Electron density confined in potentials that are consistent with the symmetry of the classical charge distribution at high magnetic field form rigid single-electron islands
that are stable against the tip potential and thus should be relatively easily resolved by the scanning probe microscopy.

\section*{Acknowledgements}
This work was supported by National Science Centre
according to decision DEC-2012/05/B/ST3/03290, and
by PL-Grid Infrastructure. Calculations were performed
in ACK -- CYFRONET -- AGH on the RackServer Zeus.


\begin{thebibliography}{1}
\bibitem{uwage} The radius of the  lowest Landau level orbit is $\sqrt{\frac{\hbar}{eB}}$; for $B=10$ T is about 8 nm.
\end{thebibliography}
\end{document}